# Flexible Cancer-Associated Chromatin Configuration (CACC) Might Be the Fundamental Reason Why Cancer Is So Difficult to Cure

Gao-De Li

Chinese Acupuncture Clinic, Liverpool, UK
Email: gaode_li@yahoo.co.uk

## Abstract

We once proposed that cell-type-associated chromatin configurations determine cell types and that cancer cell type is determined by cancer-associated chromatin configuration (CACC). In this paper, we hypothesize that flexible cell-type-associated chromatin configuration is associated with cell potency and has an advantage over inflexible one in regulating genome related activities, such as DNA replication, DNA transcription, DNA repair, and DNA mutagenesis. The reason why cancer is so difficult to treat is because CACC is flexible, which enables cancer cells not only to produce heterogeneous subclones through limited cell differentiation, but also to maximally and efficiently use genome related resources to survive environmental changes. Therefore, to beat cancer, more efforts should be made to restrict the flexibility of CACC or to change CACC so that cancer cells can be turned back to normal or become less malignant.

## Subject Areas

Cell Biology, Genomics, Molecular Biology

## 1. Introduction

Currently, it is well accepted that heterogeneity in cancers is the major reason why most cancers are so difficult to cure [1] [2]. However, the mechanism by which heterogeneous cancer cells or subclones in cancer cell populations are produced is not fully elucidated. The mainstream viewpoints blame genome instability for being the cause of cancer heterogeneity [3] [4], but we think that genome instability is only a superficial description about cancer genome, not a mechanism of how cancer heterogeneity is generated. Furthermore, we don't think that heterogeneity is the fundamental obstacle to beating cancer.

More than 30 years ago, we published a hypothesis that abnormal chromatin configuration might be associated with oncogenesis [5] [6], which has been strongly supported by research evidence [7] [8]. Recently, we published another hypothesis that a group of unknown rivet proteins might be involved in the formation of cell-type-associated chromatin configuration which determines a cell type and that cancer is the cell type that is determined by cancer-associated chromatin configuration (CACC) [9]. Based on this hypothesis, the number of rivet protein fastened-sites in the three-dimensional (3D) genome architecture determines the flexibility of cell-type-associated chromatin configurations. Compared to normal cell's cell-type-associated chromatin configuration, CACC is more flexible, which might be the fundamental reason why cancer is so difficult to cure. Detailed description about our viewpoints is presented in this paper.

## 2. Flexible CACC Might Be Associated with Cancer Cell Potency and Cancer Heterogeneity

Based on our recently published hypothesis [9], the flexibility of cell-type-associated chromatin configuration is proportional to cell potency because more flexible chromatin configurations might regulate more gene expressions, and inversely proportional to the number of rivet protein fastened-sites in the 3D genome architecture because more rivet protein fastened-sites in the 3D genome architecture mean more restrictions on the flexibility of cell-type-associated chromatin configuration, for instance, highly differentiated cells might have the largest number of rivet protein fastened-sites in their genome architectures and thus have the least flexibility in their chromatin configurations and the least cell potency, whereas stem cells might have the least number of rivet protein fastened-sites in their genome architectures and thus have the most flexibility in their chromatin configurations and the greatest cell potency. The number of rivet protein fastened-sites in cancer genome architecture is much less than those in normal somatic cell's genome architecture but a bit more than those in somatic stem cell's genome architecture, therefore, cancer cell potency is between stem cell and somatic cell, which enables cancer cell to undergo limited differentiation.

Apparently, cell differentiation is also the process during which cell potency is gradually restricted by increasing rivet protein fastened-sites in the 3D genome architectures of cells that undergo differentiation, whereas cell dedifferentiation is the process during which cell potency is gradually increased by reducing the number of rivet protein fastened-sites in the 3D genome architectures of cells that undergo dedifferentiation. Under long-term exposure to carcinogens, both somatic cells and somatic stem cells can become cancer cells, but the mechanisms by which they become cancerous might be a bit different, for instance, a cancer cell arrived from a somatic cell might be caused by abnormal dedifferentiation which involved in the reduction of the number of rivet protein fastened-sites, whereas a cancer cell arrived from somatic stem cell might be caused by abnormal differentiation which involved in the addition of the number of rivet protein fastened-sites. However, no matter whether a cancer cell comes from a somatic cell or somatic stem cell, its cell-type-associated chromatin configuration must be CACC. We think that CACC in all cancers might have the same or similar fundamental structure because all cancers have the same hallmarks, such as limitless replicative potential, evading apoptosis, tissue invasion and metastasis etc. [10].

Cell-type-specific rivet protein fastened-site patterns in the 3D genome architectures determine cell-type-associated chromatin configurations which determine cell types. In our human body there are about 200 cell types, which are required in the maintenance of a healthy human body and thus are purposely produced during evolution from unicellular organisms to multicellular organisms. Cancer cell type is not purposely produced in human body because it kills its host. The reason why human body can produce cancer cell type is because CACC that determines cancer cell type is formed by cancer-cell-type-specific rivet protein fastened-site pattern in the genome architecture, which is accidentally selected from various randomly-generated rivet protein fastened-site patterns during long-term exposure to carcinogens. Since the formation of CACC is a rare event, which does not affect the survival of multicellular-organism populations during evolution, CACC's blueprint is preserved in the genome of multicellular organisms including our humans.

Under certain conditions, a cell type can be converted into another and vice versa, which could be named as cell type transition. The underlying mechanism of cell type transition is cell-type-associated chromatin configuration transition resulting from cell-type-specific rivet protein fastened-site pattern transition [9]. Cell type transitions is common in multicellular organisms, for instance, embryonic stem cell differentiates into different somatic cells, normal somatic cells or a stem cells becomes cancer cells, and differentiated cells becomes iPS cells, all of which belong to different cell type transitions. Slight change in cell-type-associated chromatin configuration might not completely change cell type but might convert one cell type to its subtype or subclone, this type of cell type transition could be named as partial cell type transition, for example, conversion of a drug-sensitive cell type to drug-resistant cell type. Cancer heterogeneity might be caused by partial cell type transition, which is closely related to cancer cell potency. Perhaps, when microenvironment is favorable for cancer cell growth, a cancer cell type will undergo cell division to produce many identical cancer cells (clone expansion), but if there are microenvironmental changes or certain differentiation-inducing molecules (for example, certain drugs), the cancer cell type will undergo limited differentiation to produce many heterogeneous cancer cells (sub-clones) which further form heterogeneous cancer populations through clone expansion. The susceptibility of these subclones to chemotherapy, radiotherapy and immunotherapy is different, some are innately resistant to them, leading to cancer recurrence and others might be responsible for cancer metastasis. Obviously, the limited differentiation performed by cancer cells causes partial cell type transitions, which perhaps is an important defensive strategy used by cancer cells to survive environmental changes. The following schematic diagram (Figure 1) summarizes our viewpoints about cancer initiation and cancer heterogeneity formation.

Taken together, at the cellular level, it seems that the reason why most cancers are so difficult to cure is due to cancer heterogeneity, but at the genome architecture level, the reason is due to flexible CACC which enables cancer cells to produce heterogeneous subclones through limited differentiation triggered by environmental changes.

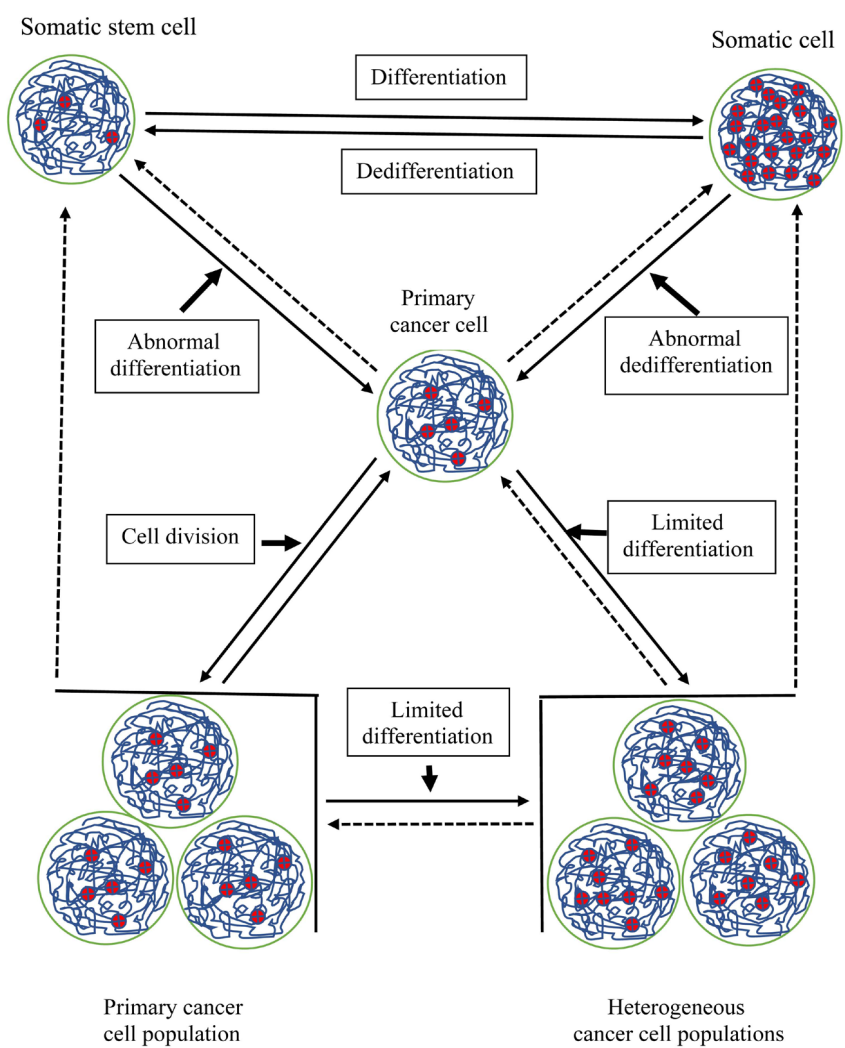

**Figure 1.** Schematic diagram of cancer initiation and cancer heterogeneity formation. Cell-type-associated chromatin configurations (blue) and rivet protein fastened-sites (red) in different cell types are shown in the diagram.

## 3. Flexible CACC Might Have an Advantage in Regulating Genome Related Activities

In the recent decades, a growing number of studies have shown that chromatin configuration or the 3D genome architecture plays an important role in regulating genome related activities, such as DNA replication, DNA transcription, and DNA repair [11]. This regulation might not be one-way, the genome related activities could also affect dynamic chromatin configuration. We once proposed that chromatin configuration and gene activity might mutually regulate each other, leading to a chain reaction-like regulation pattern during cell cycle progression [6]. Now we think that this type of regulation might also apply to cell differentiation and cell type transition.

We here propose that the flexibility of cell-type-associated chromatin configuration determines not only cell potency but also the efficiency of cell-type associated chromatin configuration in regulating genome related activities. Cell-type associated chromatin configuration sets up the framework for cell-type specific gene expression pattern and epigenetic modification pattern, and within this framework, genome related activities are regulated by dynamic genome architecture through providing suitable spatial-structure to trigger various machineries (DNA replication machinery, DNA transcription machinery, and DNA repair machinery) to work. Obviously, this type of regulation takes time, and flexible cell-type-associated chromatin configuration might make this regulation more efficient or timesaving. Therefore, compared to normal cells, cancer cell can efficiently regulate the genome related activities when they need to do so because CACC is flexible. Furthermore, flexible CACC enables cancer cells to maximally use genome related resources to survive environmental changes, for instance, cancer cells can take different pathways whenever they need to bypass the blocked pathways caused by any anticancer drugs.

We think that generation of DNA mutations, *i.e.*, DNA mutagenesis, also belongs to genome related activity. Since generation of certain gene mutations or noncoding DNA mutations involves various enzymes or proteins, it is reasonable to postulate that some DNA mutations are generated by DNA mutagenesis machinery which is also regulated by the 3D genome architecture. In another word, not all DNA mutations are completely randomly generated, some might be partially randomly generated, *i.e.*, generation of these mutations in genes or noncoding DNA regions is random, but which genes or which segments of noncoding DNA will be mutated is not random but is guided by the dynamic 3D genome architecture. For example, one anticancer drug affects a protein function, which will trigger transcription of this protein's mRNA to produce more such proteins, if the drug effects persist, the transcription activity will carry on, resulting in the formation of a transcription hotspot, which triggers the 3D genome architecture regulation, making the hotspot to move to the region in the 3D genome architecture, which is suitable for DNA mutagenesis machinery to work, or to make the hotspot regional structure suitable for DNA mutagenesis machinery to work and thus random mutations are generated within this protein gene, after natural selection some mutations will be selected so that the drug's harmful effect is diminished. Therefore, this DNA mutagenesis could be named as the 3D genome architecture guided gene mutagenesis. No doubt, preparation of regional genome structure for triggering DNA mutagenesis machinery to work takes time, for normal cells it will take quite long time because normal cell's chromatin configuration is not flexible. However, cancer cells can quickly complete this process because CACC is flexible. Some mutagenesis in noncoding DNA regions might also be regulated by the dynamic 3D genome architecture in the same way apart from the fact that the hotspot is not transcription hotspot, but any persistent blockage caused by environmental changes in the noncoding DNA regions. This DNA mutagenesis could be named as the 3D genome architecture guided noncoding DNA mutagenesis.

Drug induced DNA mutations in a single gene, or few genes might contribute to acquired drug resistance in cancer but might not reflect the whole picture of how acquired drug resistance in cancer is developed. Analysis of differential gene expression between drug-resistant and drug-sensitive cancer cells has shown that hundreds of genes are either upregulated or downregulated [12], indicating that drug resistance in cancers is not caused by a single or few mutated genes but involves wide-range gene expression changes which might be caused by slight change of cell-type-associated chromatin configuration. Therefore, it is reasonable to think that drug-resistant cancer cells might result from partial cell type transition and belong to subtypes of drug-sensitive cancer cells.

The 3D genome architecture guided gene mutagenesis and natural selection could be considered as a type of gene regulation that regulates gene product quality not gene product quantity. Clearly, in evolution, gene product quality related gene regulation is more efficient than gene product quantity related gene regulation. In addition, the 3D genome architecture guided noncoding DNA mutagenesis might also play a role in regulation of gene expression through changing regional genome structure. Due to their flexible CACC, cancer cells are prone to using these two types of gene regulation to survive environmental changes, which might be the reason why in cancer cells there are so many mutations in various genes and noncoding DNA regions [13] [14].

In conclusion, flexible CACC enables cancers to undergo limited cell differentiation to generate various subclones and to efficiently use genome related resources to survive environmental changes, which is the fundamental reason why cancer is difficult to cure.

## 4. Implications

More than 30 years ago, we proposed that abnormal chromatin configuration might cause cancer and thus turning cancer cell's abnormal chromatin configuration to normal by any means necessary will be able to turn cancer cells back to normal cells [5]. This is the first viewpoint proposed to treat disease (cancer) by manipulating chromatin configuration, which indicates a possibility of developing a novel class of drug, *i.e.*, chromatin-configuration-manipulating drugs for treating chromatin configuration related diseases.

In this paper we point out that flexible CACC is the fundamental reason why cancer is so tough to cure. Due to this reason, no matter what therapies we use in the treatment of cancers, cancer cells will be able to find way out of the harmful situation and will return eventually. Certainly, it is impossible to kill all cancer cells by currently used therapies (chemotherapy, radiotherapy and immunotherapy) unless to kill the cancer patient first. If we cannot beat cancer by cancer-cell killing strategy, we should find other ways to deal with cancer, for example, changing cancer cell behaviour through restricting the flexibility of CACC, perhaps, cancer differentiation therapy [15] is the therapy that restricts the flexibility of CACC. We can also use less toxic agent to completely or partially change CACC to elicit cell type transition or partial cell type transition so that cancer cells might become normal or less malignant and thus can live within human body like benign tumours.

As mentioned above, CACC in all cancers might have the same or similar fundamental structure which is determined by cancer-cell-type-specific rivet protein fastened-site pattern in the 3D genome architecture. The unoccupied rivet protein fastened-sites are named as rivet holes which are the structures formed by DNA and related proteins [9]. If we know the DNA sequence involved in the formation of rivet holes which are necessary for constructing the fundamental structure of CACC, we can remove or modify this DNA sequence by human embryo DNA editing so that there is no chance to form CACC in the 3D genome architecture of the baby developed from this embryo. Possibly, in this way, some babies with innate resistance to cancer will be born.

## 5. Conclusions

Humans have been fighting cancers for centuries, but at present, most cancers remain unbeatable. Currently, mainstream viewpoints blame cancer heterogeneity for being a major problem in the treatment of cancers. In this paper, we propose that the fundamental reason why most cancers are incurable is because CACC is flexible, which enable cancer cells to efficiently use genome related resources to survive environmental changes. Hence, to beat cancer, more efforts should be made to restrict the flexibility of CACC or to change CACC so that cancer cells can be turned back to normal or become less malignant.

To support our viewpoints, identification of rivet proteins is critical. We once proposed that *Plasmodium falciparum* chloroquine resistance marker protein (Pfcrmp) might be one of such rivet proteins because it contains both DNA-binding and histone-binding domains (Q968Y0_PLAFA) [9]. We believe that Pfcrmp or its homologue exists in all eukaryotic cells including human cells. Therefore, further investigation of Pfcrmp's role in the construction of the 3D genome architecture might help to unveil the mystery of CACC formation.

## Conflicts of Interest

The author declares that there is no conflict of interest regarding the publication of this paper.